\documentclass[a4paper,fleqn]{cas-sc}

\usepackage[authoryear]{natbib}
\usepackage{amsfonts,amstext,amssymb,amsmath}
\usepackage{calrsfs}
\usepackage[svgnames,dvipsnames]{xcolor}
\usepackage{multicol}
\usepackage{subfig}
\usepackage{graphicx}
\usepackage{nicefrac}
\usepackage{physics}
\usepackage{todonotes}
\usepackage{ulem}

\usepackage{tabularx}
\usepackage{booktabs}
  
\usepackage{hyperref}
\usepackage{wrapfig}
\usepackage[noabbrev,capitalise,nameinlink]{cleveref}
\hypersetup{colorlinks={true},linkcolor={blue},citecolor=blue}

\DeclareMathAlphabet{\pazocal}{OMS}{zplm}{m}{n}

\def\tsc#1{\csdef{#1}{\textsc{\lowercase{#1}}\xspace}}
\tsc{WGM}
\tsc{QE}
\tsc{EP}
\tsc{PMS}
\tsc{BEC}
\tsc{DE}
\begin{document}
\let\WriteBookmarks\relax
\def\floatpagepagefraction{1}
\def\textpagefraction{.001}

\shorttitle{Programmable wrinkling for thin circular membranes}

\shortauthors{Sairam Pamulaparthi Venkata et~al.}

\title [mode = title]{Programmable wrinkling for functionally-graded auxetic circular membranes}                      
\tnotemark[1]

\tnotetext[1]{This research project funded by the XS-META ITN: Marie Sklodowska-Curie European Actions (Grant Agreement No. 956401).}

% \tnotetext[2]{The second title footnote which is a longer text matter
%   to fill through the whole text width and overflow into
%   another line in the footnotes area of the first page.}

\author[1]{Sairam Pamulaparthi Venkata}[type=,
                        auid=000,bioid=1,
                        prefix=,
                        role=,
                        style = english,
                        orcid=0000-0002-2409-7332]


\ead{S.PamulaparthiVenkata1@universityofgalway.ie}


\credit{Writing, methodology, coding}

\affiliation[1]{organization={School of Mathematical and Statistical Sciences, University of Galway},
    addressline={University Road}, 
    city={Galway},
    postcode={H91 TK33}, 
    country={Ireland}}

\author[1]{Valentina Balbi}[
   orcid=0000-0002-7538-9490
   ]
%\fnmark[3]
\ead{vbalbi@universityofgalway.ie}
%\ead[URL]{http://valentinabalbi.weebly.com/}

\credit{Conceptualisation, writing}

\author[1,2]{Michel Destrade}[style=english, orcid=0000-0002-6266-1221]
\ead{michel.destrade@universityofgalway.ie}
% \ead[URL]{www.sayahna.org}

\affiliation[2]{organization={Key Laboratory of Soft Machines and Smart Devices of Zhejiang Province and Department of Engineering Mechanics, Zhejiang University},
    postcode={Hangzhou 310027}, 
    country={People’s Republic of China}}

\credit{Conceptualisation, writing}

\author[3]{Dino Accoto}[style=english,orcid=0000-0001-9039-5488]
%\cormark[2]
%\fnmark[1,3]
\ead{dino.accoto@kuleuven.be}
%\ead[URL]{www.stmdocs.in}

\affiliation[3]{organization=KU Leuven,
    addressline={Oude Markt 13}, 
    city={Leuven},
    % citysep={}, % Uncomment if no comma needed between city and postcode
    postcode={3000 Leuven}, 
 %   state={},
    country={Belgium}}
\credit{Conceptualisation, writing}

% Fifth author
\author[1]{Giuseppe Zurlo}[style=english,orcid=0000-0002-1438-5015]
\cormark[1]
%\fnmark[1,3]
\ead{giuseppe.zurlo@universityofgalway.ie}
\credit{Conceptualisation, methodology, writing}
%\ead[URL]{www.stmdocs.in}

\cortext[cor1]{Corresponding author}

\begin{abstract}
Materials with negative Poisson's ratio, also known as auxetic materials, display exotic properties such as  expansion in all directions under uni-axial tension. For their unique properties, these materials find a broad range of applications in robotic, structural, aerospace, and biomedical engineering. 

In this work we study the wrinkling behavior of thin and soft auxetic membranes, subjected to edge tractions. We show that spatial inhomogeneities of the Young modulus and of the Poisson ratio can be suitably tailored to produce non-trivial wrinkling patterns, with wrinkled regions that can appear, broaden, merge, and eventually disappear again, as the magnitude of applied tractions is increased monotonically. To model wrinkling in a functionally graded membrane, we employ the mathematically elegant and physically transparent tension field theory, an approximated method that we implement in commercially available software. 

Beyond unveiling the challenging technological potential to achieve non-standard wrinkling on-demand in auxetic membranes, our study also confirms the potential of using tension field theory to study, analytically and numerically, instabilities in functionally graded materials.
%, to deal with problems that are out of the reach of more accurate, but definitely more demanding, analytical and computational methods of analysis of instabilities taking place in thin membranes. 

\end{abstract}

% \includegraphics{grabs.pdf}

% \begin{highlights}
% \item Research highlights item 1
% \item Research highlights item 2
% \item Research highlights item 3
% \end{highlights}

% Keywords
\begin{keywords}
Auxetic materials \sep Thin membranes \sep Functionally graded materials \sep Hyperelasticity \sep Tension Field Theory \sep Wrinkling \sep Elastic Instabilities
\end{keywords}

\maketitle

%%%%%%%%%%%%%%%%%%%%

\section{Introduction}\label{intro}

%%%%%%%%%%%%%%%%%%%%%%

Under uniaxial tension, most isotropic materials elongate in the direction of stress and contract in the lateral directions.  The extent of deformation is governed by the value of the Poisson ratio, which, for conventional materials lies between 0 and 0.5. However, in principle, the theoretical value of Poisson's ratio for isotropic solids can range between -1 and 0.5 \citep{love1944treatise,timoshenko1983history}. 

Over the past few decades, active research has been carried out to explore materials with negative Poisson's ratio. These exotic materials are referred to as "auxetic" materials; they expand in the lateral directions when stretched longitudinally and contract in all directions when compressed. This behavior has been observed for some solids since the 1970s \citep{popereka1970ferromagnetic, milstein1979existence}. Due to their wide range of potential applications, research in auxetic materials rapidly gained significant prominence with the works of \citet{lakes1987foam}, \citet{wojciechowski1989negative}, \citet{evans1989microporous}, \citet{milton1992composite}, and \citet{lakes1993advances}. 

Progress in additive manufacturing techniques has recently enabled the fabrication of mechanical metamaterials with auxetic properties. Such metamaterials, obtained as the juxtaposition of geometric units or micro-cells, exhibit auxetic properties at the macro-scale emerging from the architecture at the micro-scale \citep{bertoldi2017flexible}. These structures are currently investigated in soft robotics, e.g. to obtain quick motions of deformable structures \citep{kaur2019toward,gorissen2020inflatable}, to produce controllable shape changes \citep{overvelde2016three}, and to develop compliant actuators \citep{lazarus2015soft}.

Nonlinear elastic membranes have been extensively used as engineering structures in aerospace and civil engineering, due to their thin, lightweight and excellent resistance under tension, with solar sails, airbags, and balloons being some of the representative examples \citep{fu2016solar}. Soft tissues such as skin and arterial cell walls also fall under this category \citep{evans2009implementation}. However, because of their almost negligible bending rigidity, membrane structures cannot sustain in-plane compressive stresses and they lose mechanical stability instantly, leading to wrinkling phenomena \citep{timoshenko20theory, roddeman1987wrinkling, cerda2002wrinkling}. 

The study of wrinkling in elastic membranes is a classical topic of mechanics. The most refined methods of analysis are based on the theory of incremental deformations \citep{haughton1978incr, haughton1978increm}, the Föppl–von Kármán theory of plates \citep{Puntel2011FvK, dymshames1973FvK}, numerical bifurcation-continuation analysis \citep{Healey2013}, reduced-order finite element membrane theory \citep{damil2010influence}. These advanced models provide detailed information on the wavelength and amplitude of wrinkles, but these are also analytically and computationally expensive. 

If the focus is put on identifying the location of wrinkled regions and the orientation of wrinkles therein, a viable alternative to the aforementioned methods is provided by tension field theory. This is a mathematically elegant and analytically treatable theory, first  proposed in the works of \citet{wagner1931flat} and \citet{reissner1938tension}. The theory relies upon the observation that thin membranes have almost negligible out-of-plane bending rigidity, and henceforth cannot sustain in-plane compressive stresses. \citet{Pipkin1986TheRE} proposed that the unilateral constraint of lack of resistance to compression could be described by introducing a  ``relaxed strain energy'', that automatically sets to zero a component of stress  whenever this would be negative in the parent energy \citep{Pipkin1986TheRE,Steigmann,Pipkin1994}. This method demonstrated its applicability in a broad range of applications involving different materials like fabrics \citep{pipkin1986continuously}, anisotropic membranes \citep{PipkinAniso}, nematic elastomer membranes \citep{DeSimone2002,cesana2015effective}, magnetoelastic membranes \citep{Reddy}, and electroelastic membranes \citep{DeTommasi2011,DeTommasi2012,GREANEY201984,KhuranaA,KhuranaB}. 

In this work, we focus our attention on annular membranes. Several works have analyzed wrinkling and buckling instabilities in circular membranes, auxetic sheets, and membranes with material inhomogeneities. For example, see the works of \citet{coman2014buckling}, \citet{lim2016large},  \citet{bonfanti2019elastic}, \citet{wang2019}, \citet{dai2021}, \citet{huang2021wrinkling}, \citet{faghfouri2022buckling}, \cite{wang2022}, and \citet{dai2022}. Although many works exist in the literature on the occurrence of wrinkling in circular membranes under different boundary conditions and by using various approaches, there is little literature on the effect of material properties such as Young's modulus and negative Poisson's ratio on the regions of wrinkling in hyperelastic auxetic membranes. This is the specific focus of the present work, which we achieve by using tension field theory. 

The paper is organized as follows. In \cref{matmodel}, we introduce the Kirchhoff strain energy material model, and derive the kinematics of deformation. Next, we recap the salient features of tension field theory and relaxed strain energy functional to derive the equilibrium equations and corresponding boundary conditions for their numerical implementation in \texttt{MATHEMATICA} \citep{Mathematica} and finite element simulation in \texttt{COMSOL} \citep{multiphysics2022}. In \cref{results}, we discuss results from \texttt{MATHEMATICA} and \texttt{COMSOL} for different selections  of material properties and applied surface traction loads. 
With some prescribed distributions of functional gradients, we find that increasing the applied traction load can lead to the formation, merging or vanishing of wrinkling regions.
Conclusions and directions for future work are presented in \cref{conclusions}.

%++++++++++++++++++++++++++++++

\section{Problem description}\label{matmodel}

%+++++++++++++++++++++++++++++++++

%%%%%%%%%%%%%%%%%%%%%%%%%%%%%%%%%%%%%%%%%
\begin{figure*}[!ht]
\centering
\includegraphics[scale=0.55]{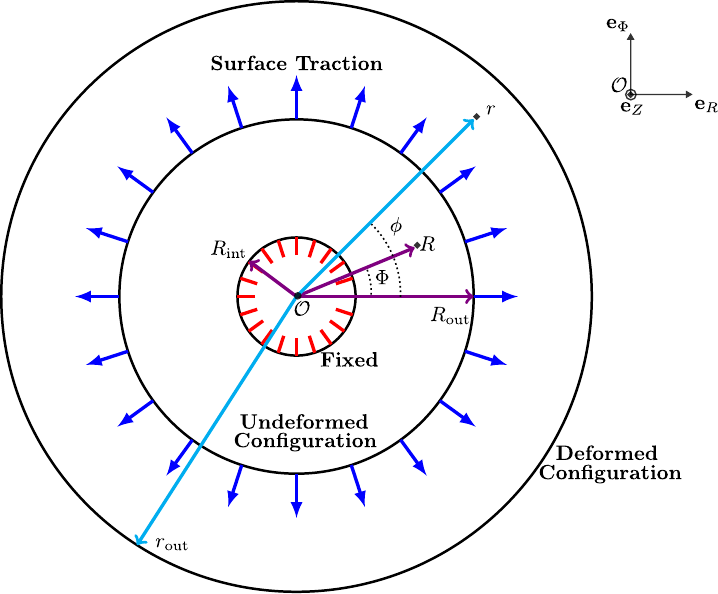}
    \caption{Undeformed and deformed configurations of a circular membrane.   The reference and current coordinates of a material point are $(R, \Phi, Z)$ and $(r, \phi, z)$, respectively, with associated basis vectors $\lbrace \mathbf{e}_{R}, \mathbf{e}_{\Phi}, \mathbf{e}_{Z} \rbrace$.
    The inner and outer radii of the undeformed and deformed membrane are $R_{\text{int}}$, $R_{\text{out}}$ and $r_{\text{int}}$, $r_{\text{out}}$, respectively.
    The inner rim of the membrane is fixed and a surface traction is applied on the outer edge.}
    \label{fig:mathematicafigure}
\end{figure*}
%%%%%%%%%%%%%%%%%%%%%%%%%%%%%%%%%%%%%%%

We consider a compressible auxetic annular disk with inner radius $R_{\text{int}}$ ($r_{\text{int}}$) and outer radius $R_{\text{out}}$ ($r_{\text{out}}$) in the reference (current) configuration, as shown in  \cref{fig:mathematicafigure}. Given its auxetic nature, the disk has a negative Poisson ratio. 
Moreover, we let the Young modulus $E$ and the Poisson ratio $\nu$ vary radially across the membrane. We identify the position of a point in the reference and current configurations with the coordinates $R,\Phi,Z$ and $r,\phi,z$, respectively. We use a single, coinciding, set of cylindrical bases both for the reference and the current configurations, namely 
 $\lbrace \mathbf{e}_{R}, \mathbf{e}_{\Phi}, \mathbf{e}_{Z} \rbrace$.
 We study the deformation and wrinkling instabilities of an annular membrane where the inner rim is fixed and a radial traction is applied to the outer rim. \par
 The deformation of the membrane is written as follows:
\begin{equation}
r = r(R), \qquad \phi = \Phi, \qquad z = z(Z).
\label{eq:1}
\end{equation}

The corresponding deformation gradient is given by:
\begin{equation}
\mathbf{F} = \begin{pmatrix}
\lambda_{R} & 0 & 0\\
0 & \lambda_{\Phi} & 0 \\
0 & 0 & \lambda_{Z}
\end{pmatrix},
\label{eq:2}
\end{equation}
where $\lambda_{R}=\nicefrac{\text{d}{r}}{\text{d}{R}}$, $\lambda_{\Phi} = \nicefrac{r}{R}$ and $\lambda_{Z}=\nicefrac{\text{d}{z}}{\text{d}{Z}}$ are the principal stretches.
We consider the 3D strain energy density (function of the three principal stretch ratios) for a compressible Kirchhoff material \citep{truesdell1960classical, truesdell2004non}, in the following form:
%\begin{equation}
%\widehat{W}^{3\text{d}}= \frac{E}{1+\nu}\left[\frac{1-\nu}{2(1-%2\nu)}I_{1}^2 - I_{2}\right],
%\label{eq:3a}
%\end{equation}
%where $E\!=\!E(R)$ and $\nu\!=\!\nu(R)$ are the Young modulus and Poisson's ratio at any radial point $(R)$ in the undeformed membrane, and $I_{1}\!=\!\trace{\mathbf{E}}$ and $I_{2}\!=\!\nicefrac{1}{2}\left[\left(\trace{\mathbf{E}}\right)^2 - \trace{(\mathbf{E}^2)}\right],$ are the two principal invariants of the Green-Lagrange strain tensor $\mathbf{E} = (\mathbf F^T\mathbf F-\mathbf I)/2$.
%We can rewrite the strain energy function in \cref{eq:3a} with respect to the principal stretches, as follows:
\begin{equation}
\begin{split}
W^{\text{3D}} = 
& \frac{E}{8(1+\nu)}\Big(\left(\lambda_{R}^2-1\right)^2 + \left(\lambda_{\Phi}^2-1\right)^2 + \left(\lambda_{Z}^2-1\right)^2 +  \frac{\nu}{1-2\nu}\left(\lambda_{R}^2 + \lambda_{\Phi}^2 + \lambda_{Z}^2 - 3\right)^2 \Big).
\end{split}
\label{eq:4}
\end{equation}

Because the thickness of the membrane in the $Z$-direction is negligible compared to its radius (for \texttt{COMSOL}, the initial thickness is taken as $H=R_\text{int}/1000$), we assume that the material is in a plane-stress state, i.e. $P_{zZ}^{\text{3D}}  = 0$ where $\mathbf{P}^{\text{3D}}  = \partial W^{\text{3D}} /{\partial \mathbf{F}}$ is the first Piola-Kirchhoff stress. Using plane-stress condition, we obtain the out-of-plane principal stretch ratio $\lambda_{Z}$ as:
\begin{equation}
\lambda_{Z} = \sqrt{\frac{1-(\lambda_{R}^2 + \lambda_{\Phi}^2-1)\nu}{1-\nu}}, \quad \nu \leq 0.
\label{eq:5}
\end{equation}

By substituting \cref{eq:5} in \cref{eq:4}, we get the strain energy function in terms of the in-plane principal stretch ratios $\lambda_{R}$ and $ \lambda_{\Phi}$, as:
\begin{equation}
\begin{split}
W = & \frac{E}{8(1-\nu^2)}\Big( \lambda_{R}^4 + \lambda_{\Phi}^4 - 2(\lambda_{R}^2 + \lambda_{\Phi}^2) + 2\nu (\lambda_{R}^2 - 1) (\lambda_{\Phi}^2-1) + 2\Big).
\end{split}
\label{eq:6}
\end{equation}

We refer to this as the \textit{membrane strain energy function} (a combination of 3D strain energy density function in \cref{eq:4} and plane-stress condition in \cref{eq:5}). Now, we can calculate the components of the first Piola-Kirchhoff stresses associated with the membrane strain energy, as:
\begin{equation}
\begin{split}
P_{11} (\equiv P_{rR}) &= \pdv{W}{\lambda_{R}} = \frac{E\lambda_{R}}{2(1-\nu^2)}\Big( \lambda_{R}^2 -1 + \left(\lambda_{\Phi}^2-1\right)\nu \Big), \\
P_{22} (\equiv P_{\phi\Phi}) &= \pdv{W}{\lambda_{\Phi}} = \frac{E\lambda_{\Phi}}{2(1-\nu^2)}\Big( \lambda_{\Phi}^2 -1 + \left(\lambda_{R}^2-1\right)\nu \Big).
\end{split}
\label{eq:7}
\end{equation}

%++++++++++++++++++++++++++++++++

\subsection{Tension Field Theory}

%++++++++++++++++++++++++++++++++

% We use Tension Field Theory \citep{pipkin1986continuously} to quantitatively study the regions of wrinkling in the membrane. 
Following \citep{Pipkin1986TheRE}, we assume that in the limit of negligible bending stiffness, wrinkles immediately appear at the outset of in-plane compressive stresses. To capture this feature, we introduce a ``relaxed strain energy density'' $W^{\pazocal{R}}$, that automatically sets a component of stress to zero, whenever this would be negative in the parent energy (membrane energy in \cref{eq:6}): 

\begin{equation}
W^{\pazocal{R}}=\begin{cases}
\begin{array}{lccc}
W\left(\lambda_{R}, \lambda_{\Phi}\right) & \text { if } & \lambda_{R} \geq \lambda_{R}^{*}\left(\lambda_{\Phi},\nu\right), & \lambda_{\Phi} \geq \lambda_{\Phi}^{*}\left(\lambda_{R},\nu\right), \\

W\left(\lambda_{R}, \lambda_{\Phi}^{*}\left(\lambda_{R}, \nu \right) \right) & \text { if } & \lambda_{R} \geq 1, &  \lambda_{\Phi}  \leq \lambda_{\Phi}^{*}\left(\lambda_{R}, \nu\right), \\

W\left(\lambda_{R}^{*}\left(\lambda_{\Phi}, \nu\right), \lambda_{\Phi}\right) \hspace{1mm} &  \text { if } & \lambda_{\Phi} \geq 1, & \lambda_{R} \leq \lambda_{R}^{*}\left(\lambda_{\Phi}, \nu\right), \\

0 \hspace{2mm} &  \text { if } &  \lambda_{R} \leq 1, &  \lambda_{\Phi}\leq 1. 
\end{array}
\end{cases}
\label{eq:8}
\end{equation}
Here, $\lambda_{i}^{*}\left(\lambda_{j},\nu\right)$ is the natural width in tension, obtained by solving $P_{ii}$ ($i = 1,2$) = 0 (no summation over $i$).
This gives us the lateral width of the membrane along the direction $i$ when the membrane is pulled along the direction $j$. In other words, if $\lambda_{\Phi} < \lambda_{\Phi}^{*}$, the membrane would be shorter than the stress-free width and hence, it would be compressed along the circumferential direction, leading to the formation of wrinkles parallel to the radial direction. 
Similarly, if $\lambda_{R} < \lambda_{R}^{*}$, wrinkles would appear parallel to the circumferential direction. 
If $\lambda_{R} > \lambda_{R}^{*}$ and $\lambda_{\Phi} > \lambda_{\Phi}^{*}$, then no wrinkles would appear in the membrane.\par
Therefore, to calculate the natural width $\lambda_{\Phi}^{*}$ when the membrane is stretched along the radial direction, we use \cref{eq:7} and we impose that: 
\begin{equation}
P_{22} =0, \quad \Rightarrow \quad \lambda_{\Phi}^{*}\left(\lambda_{R}, \nu\right) = \sqrt{1 + \nu - \lambda_{R}^2 \nu}.
\label{eq:9}
\end{equation}
Similarly, we calculate the natural width $\lambda_{R}^{*}$ when the membrane is stretched along the circumferential direction, as follows:
\begin{equation}
P_{11}  = 0, \quad \Rightarrow \quad \lambda_{R}^{*}\left(\lambda_{\Phi}, \nu\right) = \sqrt{1 + \nu - \lambda_{\Phi}^2 \nu}.
\label{eq:10}
\end{equation}
From \cref{eq:9} and \cref{eq:10}, when $\nu >0$, we observe that $\lambda_{R}, \lambda_{\Phi} < \sqrt{1+1/\nu}$ for all admissible natural widths $(\lambda_{R}^{*}, \lambda_{\Phi}^{*})$.
Hence, for a conventional Kirchhoff material ($\nu>0$), there is a threshold value of stretch, where the stress values blow to infinity and natural width values become complex. However, for an auxetic material ($-1< \nu <0$), there is no such constraint on natural widths.

%++++++++++++++++++++++++++++

\subsection{Equilibrium problem and boundary conditions}

%++++++++++++++++++++++++++++

The deformation of the membrane is governed by the balance of linear momentum $\text{Div}\mathbf{P}^N\!=\!\boldsymbol{0}$, where $\mathbf{P}^N = \lbrace \mathbf{P}^\pazocal{R}, \mathbf{P} \rbrace$ and $\text{Div}$ is a divergence operator in referential configuration. Here $\mathbf{P}^\pazocal{R}$ and $\mathbf{P}$ refer to relaxed and membrane first Piola-Kirchhoff stresses respectively. As we are looking for axially-symmetric solutions, the governing equations depend only on the radial coordinate $R$ and reduce to the following equation:
\begin{equation}
R\dv{P_{rR}^N}{R} + (P_{rR}^N - P_{\phi\Phi}^N) = 0.
\label{eq:17}
\end{equation}
We assume that the inner radius $R_{\text{int}}$ of the membrane is fixed and that a surface traction (with respect to per unit undeformed area) is applied on the outer radius $R_{\text{out}}$. The boundary conditions can then be written as follows:
\begin{equation}
\begin{split}
r\left(R_{\text{int}}\right) &= R_{\text{int}},\quad P_{rR}^N\left(R_{\text{out}}\right) = \text{Prescribed value}.
\end{split}
\label{eq:14}
\end{equation}
We solve numerically (first with \texttt{MATHEMATICA} and then with \texttt{COMSOL}) the problem in \cref{eq:17} and \cref{eq:14} to obtain the current radial coordinate $r$. Later, we use the relations in \cref{eq:2} to obtain the principal stretch ratios. Using the principal stretch ratios and \cref{eq:7}, we compute the stress components $P_{rR}^{\pazocal{R}}, P_{\phi\Phi}^{\pazocal{R}}, P_{rR}$, and $P_{\phi\Phi}$ for different spatial distributions of the material parameters (the Young modulus and the Poisson ratio) across the membrane. We then discuss how the spatial variation of these parameters affects the formation of wrinkles in the membrane.

%%%%%%%%%%%%%%%%%%%%%%%%%%%%%

\section{Numerical Results}\label{results}

%%%%%%%%%%%%%%%%%%%%%%%%%%%%%

In this section, we consider the following three scenarios. In Case 1 we assume that the Young modulus varies linearly with the radial coordinate $R$ (i.e. the radius in the reference configuration) and the Poisson ratio follows a Gaussian distribution with respect to $R$. In Case 2, we fix the Poisson ratio to a constant value and we let the Young modulus vary as a two-step function with respect to $R$. Finally, in Case 3, we take the Young modulus to vary as a step-function with respect to the radial coordinate $r$ (i.e. the radius in the deformed configuration) and the Poisson ratio to vary linearly with $r$.
%+++++++++++++++++++++++++

\subsection{Case 1: Gaussian variation of the Poisson ratio}\label{CaseI}

%+++++++++++++++++++++++++

%---------------------------------------------
% \subsubsection{Material properties}
%---------------------------------------------

When an auxetic membrane is under tension, it expands in all directions.
Due to the fixed boundary condition at the inner edge, we expect that compressive stresses might develop in that neighbourhood and possibly, wrinkles. 
To allow for that possibility, we consider the material to be highly auxetic near the inner rim. Such property can be achieved by assuming the Poisson ratio to follow a Gaussian distribution across the membrane. We take the Young modulus to vary linearly with respect to the radius $R$. Accordingly, we write $E$ and $\nu$ as follows:
\begin{equation}
\begin{split}
E\left(R\right) &= E_{\text{out}} + \Big(\frac{E_{\text{out}} - E_{\text{int}}}{R_{\text{out}} - R_{\text{int}}}\Big)(R - R_{\text{out}}), \quad \nu\left(R\right) = c_0 e^{-(c_1\left(R\right))^2},
\end{split}
\label{eq:A1a}
\end{equation}
where,
$c_1\left(R\right)\!=\!\frac{R-\text{offset}}{\text{divider}}$ and $E_{\text{out}}, E_{\text{int}}, c_0$, offset and divider are constants.

\begin{figure*}[!ht]
\centering
\includegraphics[width=0.45\textwidth]{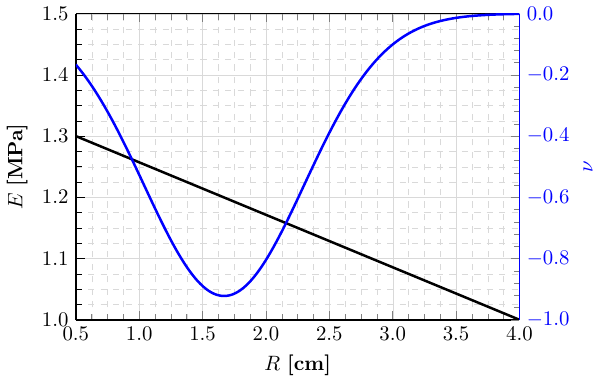}
    \caption{Case 1: Distribution of Young's modulus $E\left(R\right)$ (black line) and Poisson's ratio $\nu\left(R\right)$ (blue curve) in the membrane with respect to the referential radial coordinate $R$. The curves are obtained from \cref{eq:A1a} with the following parameters:  $E_{\text{int}}= 1.30 \hspace{1mm} \text{MPa}$, $E_{\text{out}} = 1.00 \hspace{1mm} \text{MPa}$, $c_0  = -0.9223$,
$\text{offset} = 1.670 \hspace{1mm} \text{cm}$, $\text{divider} = 0.8941 \hspace{1mm} \text{cm}$, $R_{\text{int}} = 0.5 \hspace{1mm} \text{cm}$, 
$R_{\text{out}} = 4.0 \hspace{1mm} \text{cm}$.}
    \label{fig:matpropset1}
\end{figure*}

 We found that a linearly decreasing Young modulus generates wrinkles at a lower value of applied traction load, compared to a linearly increasing Young modulus profile. So, for Case 1, we consider Young's modulus with a linearly decreasing profile and Gaussian distribution of the Poisson ratio as shown in \cref{fig:matpropset1}.

\begin{figure*}[!ht]
\centering
\includegraphics[width=0.4\textwidth]{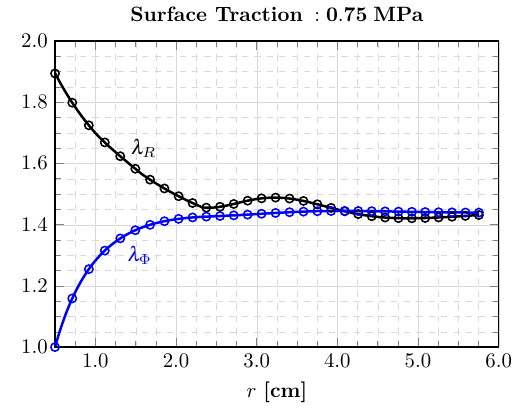}
    \caption{Case 1: Radial and circumferential components of the deformation gradient, $\lambda_{R}\left(r\right)$ and $\lambda_{\Phi}\left(r\right)$, plotted with respect to the current radial coordinate $r$ for an applied surface traction of 0.75 MPa. The solid lines and markers are results obtained from \texttt{MATHEMATICA} and \texttt{COMSOL}, respectively.}
    \label{fig:curdeformset1}
\end{figure*}
%%%%%%%%%%%%%%%%%%%%%%%%%%%%%%%%%%%%%%%

Since the material is under radial tension, the circumferential stretch $\lambda_{\Phi} \!=\!r/R$ increases with $r$. However, because the membrane is auxetic at any point, $\lambda_{\Phi} \geq 1$ implies that $\lambda_{R} \geq 1$. This behavior could be observed in \cref{fig:curdeformset1}, where for a fixed traction load, we plot the principal stretches with respect to the current radial coordinate $r$.
% We see that the plot of $F_{rR}$ is curvy, mainly due to the Poisson ratio variation shown in \cref{fig:matpropset1}. 

\begin{figure*}[!ht]
\centering
\includegraphics[width=\textwidth]{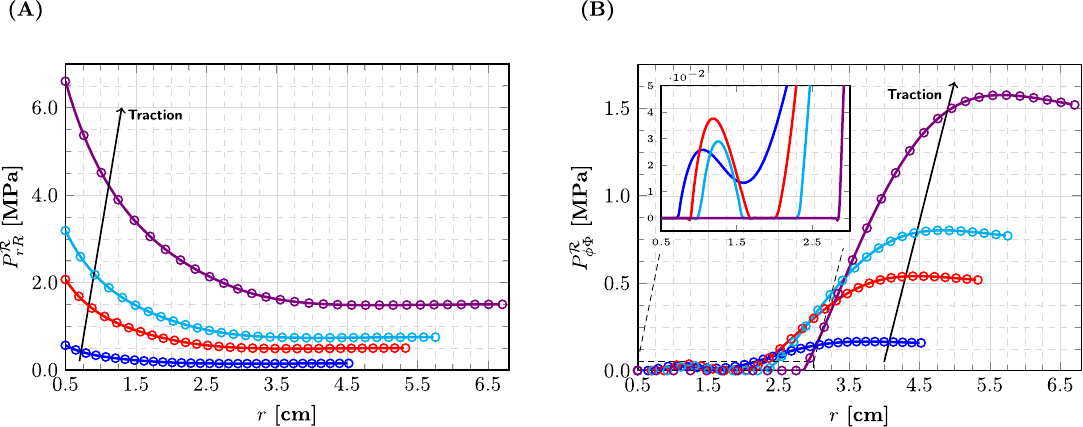}
    \caption{Case 1: The results from \texttt{MATHEMATICA} and \texttt{COMSOL} are represented by solid lines and markers, respectively. Surface traction on the outer edge of the membrane is increasing in the direction of the arrow ($0.15\hspace{1mm} \text{MPa}, 0.50 \hspace{1mm} \text{MPa}, 0.75 \hspace{1mm} \text{MPa}, 1.50 \hspace{1mm} \text{MPa}$). For increasing traction loads, radial and circumferential components of the relaxed first Piola-Kirchhoff stress, ($P^{\pazocal{R}}_{rR}$) and ($P^{\pazocal{R}}_{\phi\Phi}$), with respect to the current radial coordinate $r$ are shown in (A) and (B), respectively.}
    \label{fig:curstressset1}
\end{figure*}

The stress profiles for different traction loads are shown in \cref{fig:curstressset1}. These are obtained by solving the equilibrium equations in \cref{eq:17} and boundary conditions in \cref{eq:14}, where we have used \cref{eq:A1a} for Young's modulus and Poisson's ratio expressions. 
Note that when the circumferential stress is zero (i.e. $P_{\phi\Phi}^{\pazocal{R}} = 0$), the radial stress can be analytically deduced from the equilibrium equation \cref{eq:17} as follows:
\begin{equation}
\begin{split}
    R\dv{P^{\pazocal{R}}_{rR}}{R} + P^{\pazocal{R}}_{rR} &= 0, \qquad \Rightarrow \quad P^{\pazocal{R}}_{rR} = \frac{P_0}{R}.
\end{split}
\label{eq:19a}
\end{equation}
where the value of $P_0$ can be calculated by using the fixed boundary condition in \cref{eq:14}. We then obtain the analytical expression for the relaxed stress in the wrinkled region attached to the inner rim of the membrane as:
\begin{equation}
\begin{split}
   P^{\pazocal{R}}_{rR} = \frac{R_{\text{int}} P^{\pazocal{R}}_{rR} \left(r = R_{\text{int}}\right)}{R}.
\end{split}
\label{eq:19b}
\end{equation}

To validate our physical intuition that wrinkling occurs only along the circumferential direction, we plot \cref{fig:curstressset1}. We notice that the radial stresses are tensile and circumferential stresses contain regions of zero values in \cref{fig:curstressset1}. As we know that zero stresses correspond to the regions of wrinkling,  we conclude that wrinkling occurs only along the circumferential direction, thus meeting our expectations.

%%%%%%%%%%%%%%%%%%%%%%%%%%%%%%%%%%%%%%%%%
%%%%%%%%%%%%%%%%%%%%%%%%%%%%%%%%%%%%%%%

From the zoomed inset in \cref{fig:curstressset1}B, we observe that for all traction loads considered (0.15 MPa to 1.50 MPa), there is a region near the inner edge where the circumferential stress is zero. Moreover, in this region, the radial stress computed from \texttt{MATHEMATICA} and \texttt{COMSOL} matches with the analytical solution in \cref{eq:19b}, although, for brevity, this particular result is not shown here.

In \cref{fig:curstressset1}B, for the traction load  0.15 MPa (blue curve), there is only one region, at the inner edge, where the circumferential stress is zero. As we increase the traction load to 0.50 MPa (red curve), a new region emerges.
Then both regions start to widen as we increase the traction load further (1.00 MPa, cyan curve), until they merge into one large region, see (violet curve) for the traction load 1.50 MPa. 
For higher traction loads (up to 12.0 MPa), we find that the wrinkling region slightly grows  without the formation of new regions.

\begin{figure*}[!ht]
\centering
\includegraphics[scale=0.5]{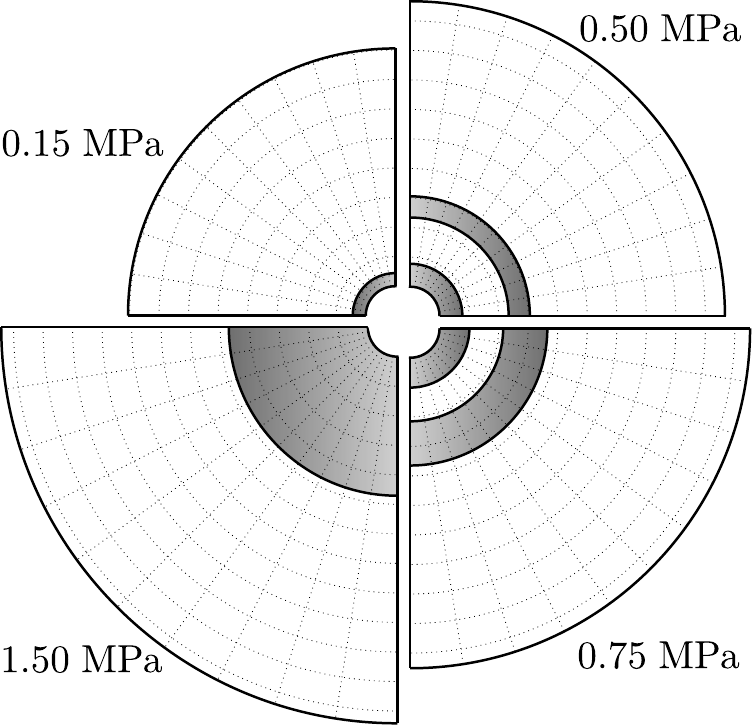}
    \caption{Case 1: Distribution of wrinkling profiles with an increase in the traction load from the left top in clockwise direction. Each quadrant refers to a different applied traction load (0.15 MPa to 1.50 MPa). Thanks to axisymmetry, only quadrants of the circular membrane are shown here. In the grey (wrinkled) regions, the circumferential stress $P^{\pazocal{R}}_{\phi\Phi}$ is zero.}
    \label{fig:curwrinklset1}
\end{figure*}

To highlight this feature, we plot the quadrants of the deformed circular membrane for each applied traction load (taking advantage of the axisymmetry in the problem) in \cref{fig:curwrinklset1}, from 0.15 MPa (left top case) to 1.50 MPa (left bottom case) in the clockwise direction. Dark regions correspond to the regions of wrinkling. The values of wrinkling regions are reported in {\color{blue}{Table S1}} (SI).

\subsection{Case 2: Two-step variation of Young's modulus}\label{CaseII}

In this section, to obtain three regions of wrinkling we fix the Poisson ratio at a constant value and we take the Young modulus to vary as a two-step function with respect to the radius $R$, as follows:
\begin{equation}
\begin{split}
E\left(R\right) &= E_0\Biggl(\frac{e^{2E_1\left(R\right)}-1}{e^{2E_1\left(R\right)}+1}\Biggl) -E_0\Biggl(\frac{1.9e^{2E_2\left(R\right)}-1}{e^{2E_2\left(R\right)}+1}\Biggl) + E_0\Biggl(\frac{e^{2E_3\left(R\right)}-1}{e^{2E_3\left(R\right)}+1}\Biggl)
-E_0\Biggl(\frac{1.2e^{2E_4\left(R\right)}-1}{e^{2E_4\left(R\right)}+1}\Biggl)+E_{\text{res}}, \\
%\nu[R] &= -0.3,
\end{split}
\label{eq:A2a}
\end{equation}
where, $E_i\left(R\right)\!=\!\frac{R-\text{offset}_i}{\text{divider}_1}$ and $E_0, E_{\text{res}}$, $\text{divider}_1$ and $\text{offset}_i$, with $i\!=\!\{1,\dots,4\}$, are constants.

\begin{figure*}[!ht]
\centering
\includegraphics[width=0.5\textwidth]{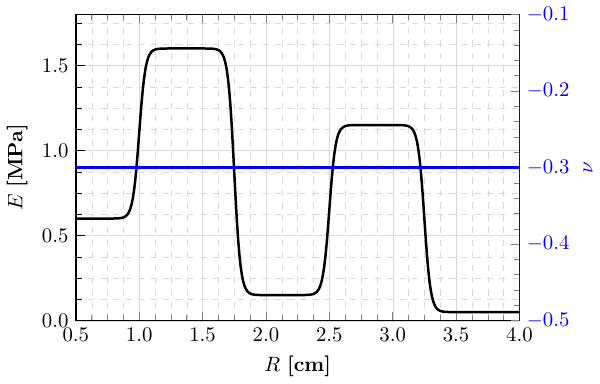}
    \caption{Case 2: Distribution of Young's modulus $E\left(R\right)$ (black curve) and Poisson's ratio $\nu\left(R\right)$ (blue line) in the membrane with respect to the referential radial coordinate $R$. The plot for the Young modulus is obtained by using \cref{eq:A2a}. The parameters are: $E_{\text{res}} = 0.60 \hspace{1mm} \text{MPa}$, $E_0= 0.50 \hspace{1mm} \text{MPa}$,
$\text{offset}_1 = 1.00 \hspace{1mm} \text{cm}$, $\text{offset}_2 = 1.75 \hspace{1mm} \text{cm}$,
$\text{offset}_3 = 2.50 \hspace{1mm} \text{cm}$, $\text{offset}_4 = 3.25 \hspace{1mm} \text{cm}$,
$R_{\text{int}} = 0.50 \hspace{1mm} \text{cm}$, $R_{\text{out}} = 4.00 \hspace{1mm} \text{cm}$,
$\text{divider}_1 = 0.05 \hspace{1mm} \text{cm}$.}
    \label{fig:matpropset3}
\end{figure*}

\begin{figure*}[!ht]
\centering
\includegraphics[width=0.9\textwidth]{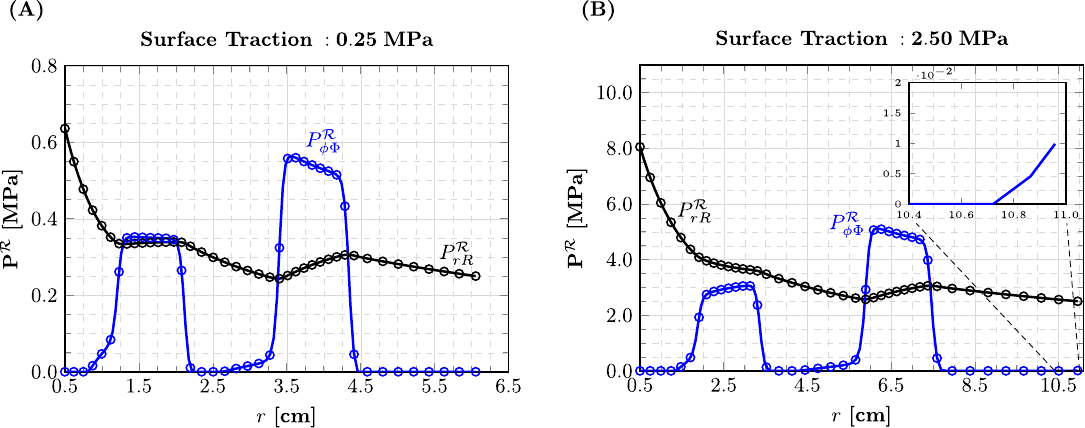}
    \caption{Case 2: Relaxed first Piola-Kirchhoff stress ($\mathbf{P}^{\pazocal{R}}$) components  with respect to the current radial coordinate ($r$) for different traction loads: (A) Traction load: 0.25 MPa, (B) Traction load: 2.50 MPa. The solid lines and markers are results obtained from \texttt{MATHEMATICA} and \texttt{COMSOL} respectively. The black solid line and black markers represent the radial component of relaxed first Piola-Kirchhoff stress ($P^{\pazocal{R}}_{rR}$),  while the blue solid line and blue markers represent the circumferential component of relaxed first Piola-Kirchhoff stress ($P^{\pazocal{R}}_{\phi\Phi}$).}
    \label{fig:curstressset3}
\end{figure*}

\begin{figure*}[!ht]
\centering
\includegraphics[scale=0.4]{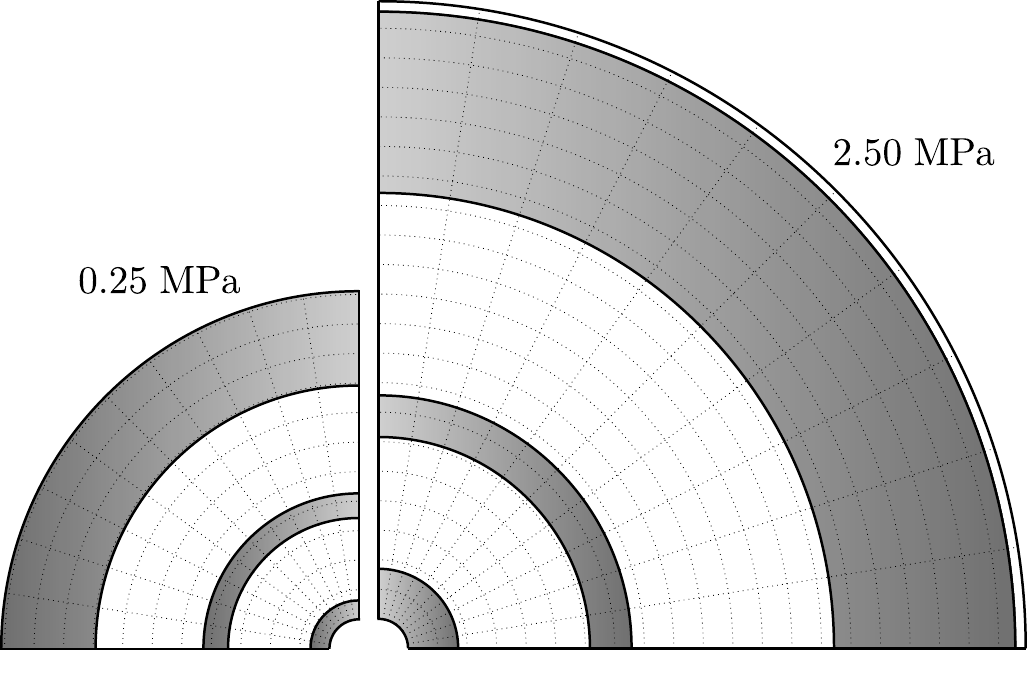}
    \caption{Case 2: Distribution of wrinkling profiles with increasing traction load from left (0.25 MPa) to right (2.50 MPa). Each quadrant refers to a different applied traction load. At the lower load, there are three wrinkled regions (grey) and two unwrinkled regions (white). At the higher load, a third unwrinkled region emerges near the outer rim.}
    \label{fig:curwrinklset3}
\end{figure*}

The profile of Young's modulus according to \cref{eq:A2a} and constant Poisson's ratio are shown in \cref{fig:matpropset3}. For Case 2, we increase the traction load from 0.25 MPa to 15 MPa and plot two representative stress plots in \cref{fig:curstressset3}. Also, the corresponding wrinkling profiles are plotted in \cref{fig:curwrinklset3}. 

For low values of traction, we find three wrinkling regions where the circumferential stress is zero (\cref{fig:curstressset3}), i.e. three regions where wrinkles appear along the circumferential direction (\cref{fig:curwrinklset3}). 
For a traction load of 0.25 MPa, we see that there is one wrinkling region close to the inner edge, one region close to the outer edge, and an intermediate region. 
Then, as we increase the traction load, the inner wrinkled region grows, while an interesting behavior is observed for the outer wrinkled region. 
Hence, when the traction is set to 2.50 MPa, we observe that a small region close to the outer edge starts to ``unwrinkle''. 
This is clearly seen from the non-zero circumferential stress in the zoomed inset plot of \cref{fig:curstressset3}B and from a thin white region occurring near the outer edge in \cref{fig:curwrinklset3}. 
By looking at the stresses in the current configuration (see \cref{fig:curstressset3}), we notice that the intermediate wrinkling region has shifted its location away from the inner fixed rim of the membrane as the traction load increased. 
The geometry of the wrinkling regions and the deformed radial coordinate are tabulated in {\color{blue}{Table S2}} (SI) for 0.25 MPa and 2.50 MPa traction loads.

%+++++++++++++++++++++++++++

\subsection{Case 3: Young's modulus and Poisson's ratio vary with current radius $r$}\label{CaseIII}

%++++++++++++++++++++++++++

Finally, we assume that the material properties vary with respect to the current radial coordinate $r$, in contrast to Cases 1 and 2, where they depend on $R$. 
Our rationale is that auxetic microstructures can change shape during the deformation and hence the effective Young modulus and effective Poisson ratio may vary with the deformation. 

Similar to Case 2, we are interested in capturing two wrinkling regions in the material; thus we choose a step function in terms of $r$ for the Young modulus while the Poisson ratio varies linearly with $r$. 
For Case 3, the Young modulus and the Poisson ratio depend on $r$, which is unknown, and are therefore unknown variables in the solution procedure, as opposed to Case 1 and 2, where the material parameters are functions of $R$, and therefore known quantities.

\begin{figure*}[!ht]
\centering
\includegraphics[width=0.9\textwidth]{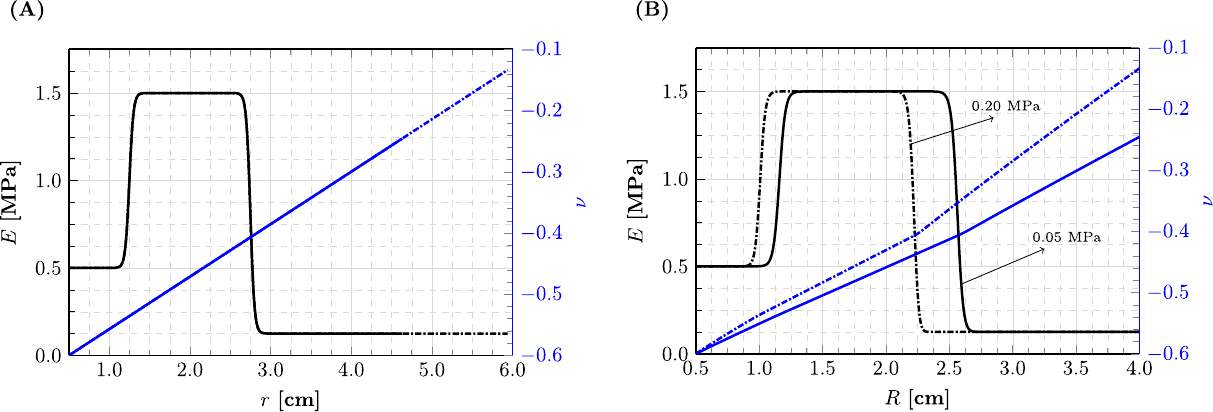}
    \caption{Case 3: Distribution of Young's modulus $E\left(r\right)$ and Poisson's ratio $\nu\left(r\right)$ in the membrane. (A) With respect to the current radial coordinate $r$;  (B) With respect to the referential radial coordinate $R$. The solid lines and dash-dotted lines correspond to traction loads of 0.05 MPa and 0.20 MPa, respectively. Once the current radial co-ordinate $r$ is obtained by solving \cref{eq:17} and \cref{eq:14}, the material property curves are plotted using \cref{eq:A3a} with the following parameters: $E_{\text{res}}^{*} = E_0^{*} = 0.50 \hspace{1mm} \text{MPa}$, $\text{offset}_a = 1.25 \hspace{1mm} \text{cm}$, $\text{divider}_a = 0.05 \hspace{1mm} \text{cm}$, $\text{offset}_b = 2.75 \hspace{1mm} \text{cm}$, $R_{\text{int}} = 0.5 \hspace{1mm} \text{cm}$,  $R_{\text{out}} = 4.0 \hspace{1mm} \text{cm}$, $\nu_{\text{out}} = -0.3$, $\nu_{\text{int}} = -0.6.$}
    \label{fig:combmatpropsmpset2}
\end{figure*}

Once we incorporate the relation between current and referential radial coordinates, we can visualize the Young modulus and the Poisson ratio as functions of the referential radial coordinate $R$, and current radial coordinate $r$, as shown in \cref{fig:combmatpropsmpset2}. As the deformation of the circular membrane is different for each applied traction load, we observe that the material properties also change with the prescribed traction load, see \cref{fig:combmatpropsmpset2}. 
The variations of Young's modulus and Poisson's ratio with respect to $r$ are given by:

\begin{equation}
\begin{split}
E\left(r\right) &= E_{\text{res}}^{*} + E_0^{*}\Biggl(\frac{e^{2E_a\left(r\right)}-1}{e^{2E_a\left(r\right)}+1}\Biggl) -E_0^{*}\Biggl(\frac{1.75e^{2E_b\left(r\right)}-1}{e^{2E_b\left(r\right)}+1}\Biggl), \qquad \nu\left(r\right) = \nu_{\text{out}} + \Big(\frac{\nu_{\text{out}} - \nu_{\text{int}}}{R_{\text{out}} - R_{\text{int}}}\Big)(r - R_{\text{out}}),
\label{eq:A3a}
\end{split}
\end{equation}
where, $E_i\left(r\right)\!=\!\frac{r-\text{offset}_i}{\text{divider}_a}$ and $E_0, E_{\text{res}},\nu_{\text{int}},\nu_{\text{out}}$, $\text{divider}_a$ and $\text{offset}_i$, with $i\!=\!\{a,b\}$, are constants.

%%%%%%%%%%%%%%%%%%%%%%%%%%%%%%%%%%%%%%%

\begin{figure*}[!ht]
\centering
\includegraphics[width=0.9\textwidth]{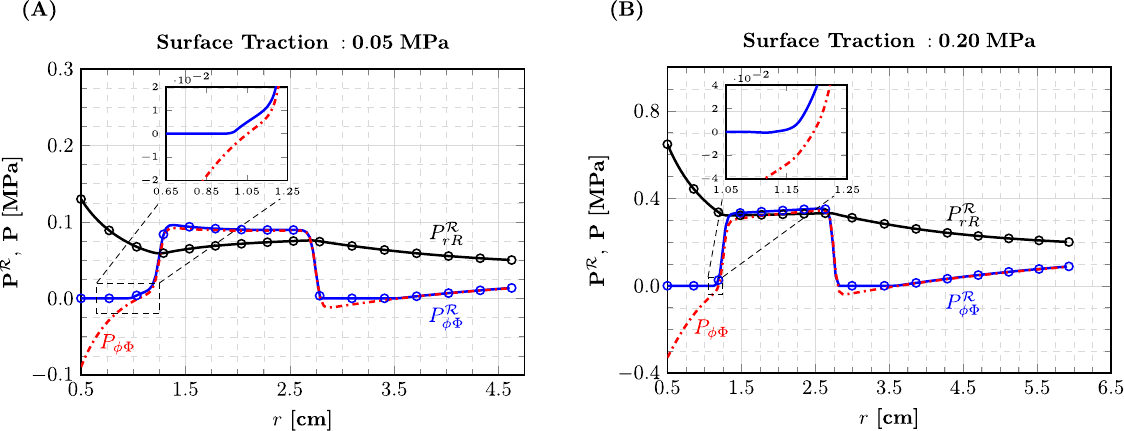}
    \caption{Case 3: Relaxed (Membrane) first Piola-Kirchhoff stress  $\mathbf{P}^{\pazocal{R}}$ ($\mathbf{P}$) components with respect to the current radial coordinate $r$ for different traction loads: (A) Traction load: 0.05 MPa, (B) Traction load: 0.20 MPa. The solid (and dash-dotted) lines and markers are results obtained from \texttt{MATHEMATICA} and \texttt{COMSOL}, respectively. The black solid line and black markers represent the radial component of relaxed first Piola-Kirchhoff stress ($P^{\pazocal{R}}_{rR}$), blue solid line and blue markers represent the circumferential component of relaxed first Piola-Kirchhoff stress ($P^{\pazocal{R}}_{\phi\Phi}$), and blue dash-dotted line represents the circumferential component of membrane first Piola-Kirchhoff stress ($P_{\phi\Phi}$).}
    \label{fig:curstressmpset2}
\end{figure*}

We solve the equilibrium equations in \cref{eq:17} and boundary conditions in \cref{eq:14}, to compute the relaxed first Piola-Kirchhoff stress components $P^{\pazocal{R}}_{rR}$ and $P^{\pazocal{R}}_{\phi\Phi}$. Furthermore, we calculate the stresses $P_{rR}$ and $P_{\phi\Phi}$ for the membrane strain energy. These are plotted in \cref{fig:curstressmpset2} for the material function in \cref{eq:A3a}.

\begin{figure}
\centering
\includegraphics[scale=0.6]{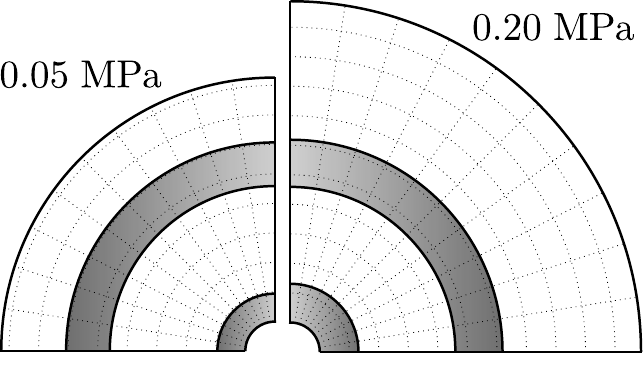}
    \caption{Case 3: Distribution of wrinkling profiles with an increase in the traction load from left (0.05 MPa) to right (0.20 MPa). Each quadrant refers to a different applied traction load.}
    \label{fig:curwrinklmpset2}
\end{figure}

We observe two wrinkling regions appearing as we increase the applied traction load. We plot the corresponding stress profiles in \cref{fig:curstressmpset2} and the corresponding wrinkling profiles in \cref{fig:curwrinklmpset2}. 

As opposed to Cases 1 and 2, although the inner wrinkling region grows in size with increasing traction load, there is hardly any growth in the other wrinkling region. This can be clearly seen from the wrinkling plot in \cref{fig:curwrinklmpset2}. The values of wrinkling regions and current radial coordinates of the membrane are tabulated in {\color{blue}{Table S3}} (SI).

Since the Poisson ratio is deformation-dependent, we remark that at higher traction loads ($ \geq 0.48$ MPa), there will be regions in the membrane with positive Poisson's ratio. As mentioned in \cref{eq:10}, natural width values for a conventional Kirchhoff material ($\nu > 0$) are not physically admissible at higher stretch ratios because there is a threshold stretch value where the stresses blow to infinity. So we restrict our attention here to negative Poisson's ratio and consider traction loads lower than 0.48 MPa. 
For representative purposes, we show the wrinkling regions of the material for traction loads of 0.05 MPa and 0.20 MPa in {\color{blue}{Table S3}} (SI). 

In \cref{fig:curstressmpset2}, we compare the circumferential stresses obtained from relaxed and membrane strain energy functions. 
The plots show that it is not the entire region under compressive stresses obtained from the unrelaxed strain energy that contributes to wrinkling, and it reinforces the importance of the relaxed strain energy function. 
We observe that the inner region of zero circumferential stresses obtained from the relaxed strain energy function is only a subset of the inner region of compressive stresses obtained from the membrane strain energy function, although the difference is not large for this case.

%%%%%%%%%%%%%%%%%%%%%%

\section{Conclusions}\label{conclusions}

%%%%%%%%%%%%%%%%%%%%%%%%%

We studied the effect of varying material properties such as Young's modulus and Poisson's ratio on wrinkling instability in auxetic circular membranes using tension field theory. 
We solved the equilibrium equations in \cref{eq:17} and boundary conditions in \cref{eq:14} numerically with \texttt{MATHEMATICA} and verified the results with finite element simulations in \texttt{COMSOL}. 
We assumed that the material properties are functions of the radial coordinates (in turn, referential $R$ and current $r$). 
We also discussed the importance of using the relaxed strain energy in comparison to the membrane strain energy for accurately capturing the wrinkling regions. 

For the purpose of this study, we considered three scenarios associated with different spatial variations of the Young modulus and the Poisson ratio. We concluded that:
\begin{enumerate}
\item We can obtain regions of wrinkling at  desired locations in the membrane by prescribing appropriate variations in material properties. 
\item With increasing applied traction load:
\begin{enumerate}
    \item We can find new wrinkled regions emerging, which can merge with the earlier existing ones;
    \item Regions of wrinkling can grow in place;
    \item The size of wrinkling regions can  remain unchanged;
    \item Unwrinkling of existing wrinkling regions can  occur.
\end{enumerate}
\end{enumerate}

The method we proposed here can be generalized to other geometries, such as rectangular membranes, and the effect of dimensions (size effects) on wrinkling instability using tension field theory can be further explored. 
The results presented here provide a proof of concept. 
Based on the insights from this study, the results can be further refined 
with more sophisticated numerical techniques to account for bending and shear stiffnesses and obtain a fine-scale description of the wrinkles, including amplitude and wavelength.

Experimentally, the fabrication of membranes with graded properties to be used in wrinkling experiments is challenging. 
With recent advancements in additive and subtractive manufacturing techniques, functionally-graded materials with complex designs can be manufactured, especially structures involving step functions of material properties and with a pointwise variable Young modulus \citep{naebe2016functionally}. 
To obtain negative and/or variable Poisson moduli, the mechanical metamaterial approach can be used \citep{ren2018auxetic}. This  introduces a further dimensional scale, for which it will be necessary on the one hand, to work with homogenization techniques, on the other hand, to adapt the model proposed here to improve its predictive power in relation to the specific fabrication technique selected, so that it can become an agile tool to support designers.
Finally, the displacement field can be measured with digital image correlation (DIC) techniques \citep{chen2023meshfree}.

This study lays the foundations for the development of analytical tools that can assist designers in the construction of morphing surfaces. These surfaces, which have the ability to assume desired 3D shapes, have numerous potential applications in Engineering. 
Here we elaborate on three examples.

In the field of Wearable Robotics, it is important to ensure the safety and comfort of the interaction surfaces, i.e. of the surfaces that connect the robot to the human body \citep{babivc2021challenges}. 
%Safety and comfort often require no sliding between the interaction surfaces and human skin. 
Traditional interaction surfaces are unable to accommodate the wrinkling exhibited by the skin during flexion-extension of the body joints. We believe that the development of surfaces capable of exhibiting programmed wrinkling can overcome this limitation, making possible the fabrication of interaction surfaces that are safer for the skin (lower risk of abrasion) and more comfortable for the user.

In Tissue Engineering, it is often necessary to create 3D scaffolds with appropriate porosity and biocompatibility properties to allow cell engraftment. Currently, 3D printing techniques are being used for the development of such scaffolds. The possibility of producing 3D geometries through controlled wrinkling would make it possible to print flat membranes which would assume their final 3D shape upon application (or removal) of stress. 
This would facilitate the deposition of cells \citep{zhao2018programmed}, and would also represent savings in terms of printing time and amount of biomaterial used.

In Aerospace Engineering, and in particular in the development of small battery-operated unmanned aerial vehicles (UAV), the need to increase aerodynamic efficiency to ensure good battery life is increasingly felt, especially in light of the ever more stringent requirements in terms of maneuverability. The possibility of changing the geometry of an aerofoil in a simple way, for example through traction, could allow the improvement of the aerodynamic efficiency of UAVs, with positive implications on energy autonomy  \citep{zahoor2020preliminary}.
\color{black}

%%%%%%%%%%%%%%%%%%%%%%%%%%%%%%%%%%
\printcredits

\section*{Declaration of Competing Interest}
The authors declare that they have no known competing financial interests or personal relationships that could have appeared to influence the work reported in this paper.

\section*{Acknowledgments}
\begin{figure*}[!ht]
\centering
\includegraphics[width=0.15\textwidth]{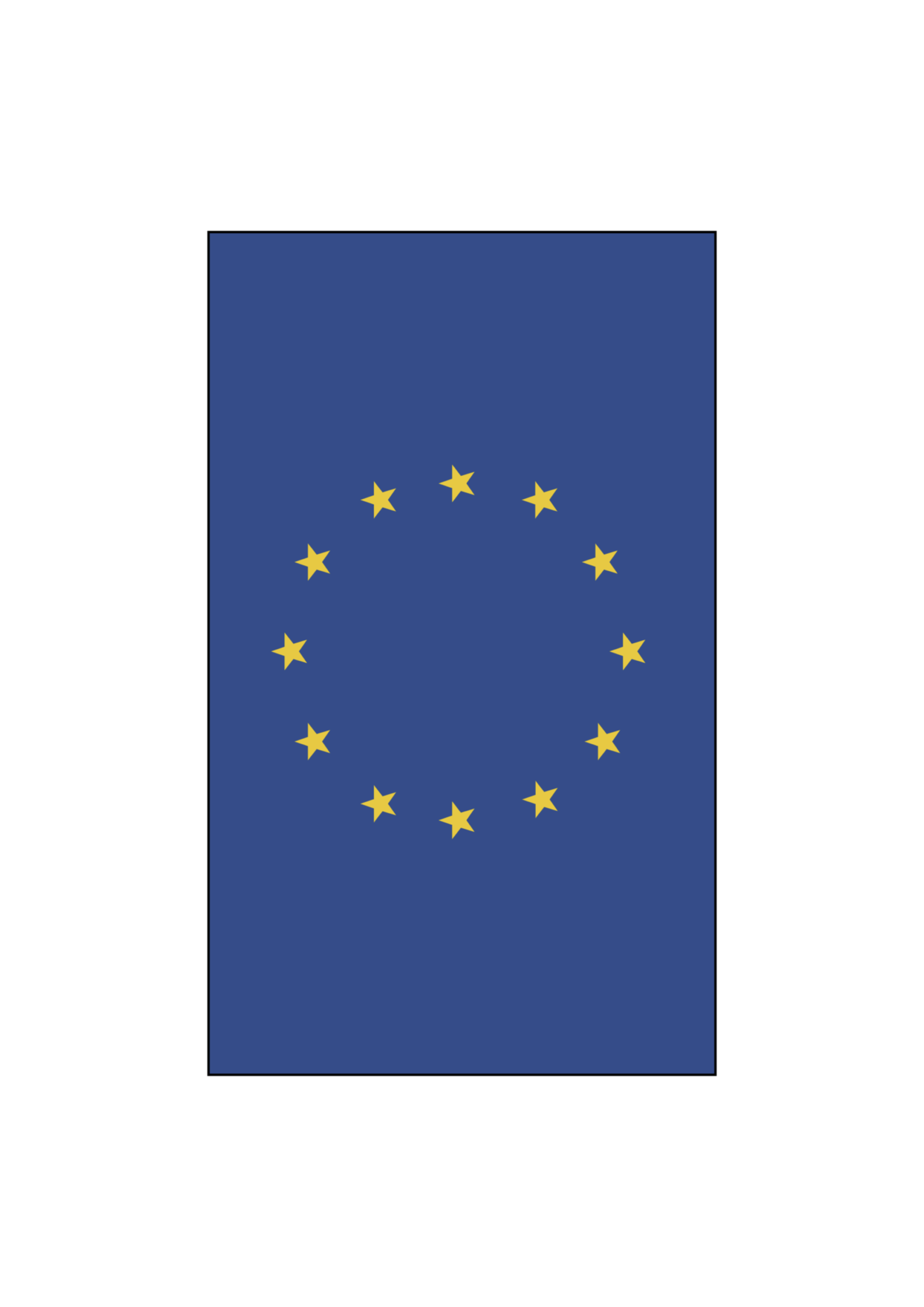}
\end{figure*}
This project has received funding from the European Union’s Horizon 2020 research and innovation programme under the Marie Skłodowska-Curie Grant Agreement No. 956401. GZ gratefully acknowledges the support of GNFM (Gruppo Nazionale di Fisica Matematica) of the INdAM F.Severi.  We thank the anonymous referees for valuable suggestions.\color{black}
\section*{Supplementary data}
The Supplementary material is attached to this article.

%\bibliographystyle{cas-model2-names}
\bibliographystyle{elsarticle-harv-doichange}
\bibliography{cas-refs}

\newpage

{\Huge \textbf{Supplementary Information}}                    

\bigskip

\renewcommand*{\thesection}{S\arabic{section}}
\setcounter{section}{0}
\renewcommand{\theequation}{S.\arabic{equation}}
% reset the counter
\setcounter{equation}{0}
\renewcommand*{\thefigure}{S\arabic{figure}}
\setcounter{figure}{0}
\renewcommand*{\thetable}{S\arabic{table}}
\setcounter{table}{0}

%\appendix

%%%%%%%%%%%%%%%%%%
%-------------------------------------------------
\section{Wolfram Mathematica implementation}
%--------------------------------------------------

By combining all the cases in {\color{blue}{Equation (7)}} (Main file), the relaxed strain energy density functional can be re-written as
\begin{subequations}
\begin{equation}
\begin{split}
W^{\pazocal{R}} &= \Big(\textsc{Heavy}_{1}\left(\lambda_{R}, \lambda_{\Phi}, \nu\right)\Big) W\big(\lambda_{R},\lambda_{\Phi}, E, \nu\big) + \Big(\textsc{Heavy}_{2}\left(\lambda_{R}, \lambda_{\Phi}, \nu\right)\Big) W\big(\lambda_{R},\lambda_{\Phi}^{*}\left(\lambda_{R}, \nu\right), E, \nu\big) \\&+ \Big(\textsc{Heavy}_{3}\left(\lambda_{R}, \lambda_{\Phi}, \nu\right) \Big) W\big(\lambda_{R}^{*}\left(\lambda_{\Phi}, \nu\right),\lambda_{\Phi}, E, \nu\big),
\end{split}
\label{eq:11a}
\end{equation}
where 
\begin{equation}
\begin{split}
&\textsc{Heavy}\left(\alpha\right) = \frac{1}{1 + e^{-\beta\alpha}}, \quad \beta = 3000,\\
&\textsc{Heavy}_{1}\left(\lambda_{R}, \lambda_{\Phi}, \nu\right) = \Big(\textsc{Heavy}\big(\lambda_{\Phi} - \lambda_{\Phi}^{*}\left(\lambda_{R}, \nu\right) \big)\Big)\Big(\textsc{Heavy}\big(\lambda_{R} - \lambda_{R}^{*}\left(\lambda_{\Phi}, \nu\right) \big)\Big), \\
&\textsc{Heavy}_{2}\left(\lambda_{R}, \lambda_{\Phi}, \nu\right) = \Big(\textsc{Heavy}\left(\lambda_{R} - 1\right)\Big)\Big(\textsc{Heavy}\big(\lambda_{\Phi}^{*}\left(\lambda_{R}, \nu\right) - \lambda_{\Phi} \big)\Big), \\
&\textsc{Heavy}_{3}\left(\lambda_{R}, \lambda_{\Phi}, \nu\right) = 
\Big(\textsc{Heavy}\left(\lambda_{\Phi} - 1\right)\Big)\Big(\textsc{Heavy}\big(\lambda_{R}^{*}\left(\lambda_{\Phi}, \nu\right) - \lambda_{R} \big)\Big).
\end{split}
\label{eq:11b}
\end{equation}
Here, $\textsc{Heavy}\left(\cdot\right)$ is the smooth Heaviside function with origin as its center. The higher the value of $\beta$ is, the closer $\textsc{Heavy}\left(\cdot\right)$ is to a step function. The value of $\beta$ is assumed to be 3000 for our analysis.

When wrinkles form only along direction 2, the relaxed energy in the wrinkling region becomes 
\begin{equation}
\begin{split}
W^{\pazocal{R}}\Big(\lambda_{R}, \lambda_{\Phi}^{*}\left(\lambda_{R}, \nu\right), E, \nu\Big) &= \frac{E}{8}\left(\lambda_{R}^2-1\right)^2.
\end{split}
\label{eq:11c}
\end{equation}
Similarly, when wrinkles form only along direction 1, the relaxed energy in the wrinkling region becomes 
\begin{equation}
\begin{split}
W^{\pazocal{R}}\Big(\lambda_{R}^{*}\left(\lambda_{\Phi}, \nu\right),\lambda_{\Phi}, E, \nu\Big) &= \frac{E}{8}\left(\lambda_{\Phi}^2-1\right)^2.
\end{split}
\label{eq:11d}
\end{equation}
\end{subequations}

Now we can combine {\color{blue}{Equation (5)}} (Main file), and \cref{eq:11a} to \cref{eq:11d} to get the full form of relaxed strain energy density functional as
\begin{equation}
\begin{split}
&W^{\pazocal{R}}\Big(\lambda_{R}, \lambda_{\Phi}, E, \nu\Big) = \frac{E\left(2-2 \lambda_{R}^{2}+\lambda_{R}^{4}-2 \lambda_{\Phi}^{2}+\lambda_{\Phi}^{4}+2\left(-1+\lambda_{R}^{2}\right)\left(-1+\lambda_{\Phi}^{2}\right) v\right)}{8\left(1+e^{-3000\left(\lambda_{\Phi}-\sqrt{1+{\nu}-\lambda_{R}^{2} {\nu}}\right)}\right)\left(1+e^{-3000\left(\lambda_{R}-\sqrt{1+{\nu}-\lambda_{\Phi}^{2} {\nu}}\right)}\right)\left(1-{\nu}^{2}\right)} \\&
+ \frac{E\left(\lambda_{R}^2-1\right)^2}{8\left(1+e^{-3000(-1+\lambda_{R})}\right)\left(1+e^{-3000\left(-\lambda_{\Phi}+\sqrt{1+\nu-\lambda_{R}^{2} \nu}\right)}\right)} 
+ \frac{E\left(\lambda_{\Phi}^2-1\right)^2}{8\left(1+e^{-3000(-1+\lambda_{\Phi})}\right)\left(1+e^{-3000\left(-\lambda_{R}+\sqrt{1+\nu-\lambda_{\Phi}^{2} \nu}\right)}\right)}.
\end{split}
\label{eq:12}
\end{equation}

%++++++++++++++++++++++++++++++++++

\section{COMSOL implementation}

%+++++++++++++++++++++++++++++++++++

%%%%%%%%%%%%%%%%%%%%%%%%%%%%%%%%%%%%%%%%%
\begin{figure}[ht]
\centering
\includegraphics[width=0.4\textwidth]{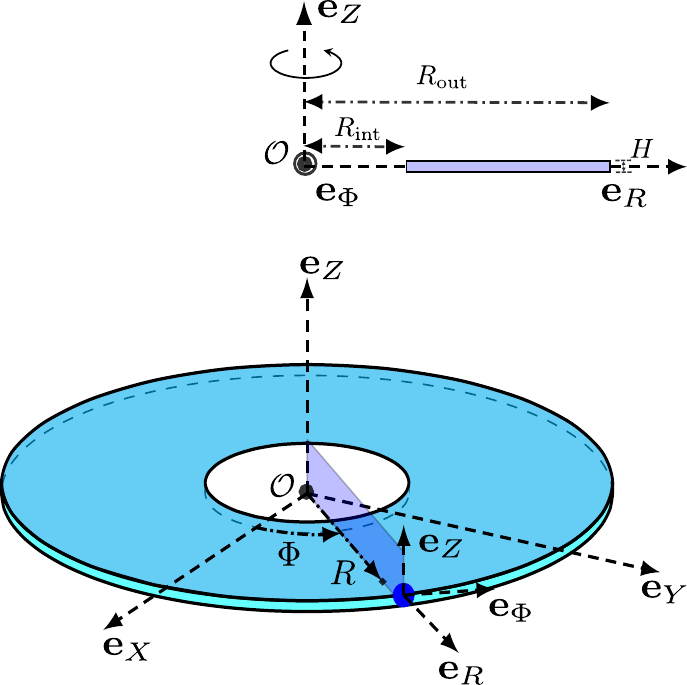}
    \caption{Revolution in the $R-Z$ plane of a thin 2d rectangular strip (thickness $H$, inner radius  $R_{\text{int}}$ and outer radius $R_{\text{out}}$ in reference configuration) to create an annular strip in \texttt{COMSOL}. }
    \label{fig:comsolfigure}
\end{figure}
%%%%%%%%%%%%%%%%%%%%%%%%%%%%%%%%%%%%%%%

To implement tension field theory in \texttt{COMSOL}, we considered our axisymmetric circular disk as a revolution geometry obtained from a 2D thin rectangular strip in the $(R,Z)$ plane with height $H (\approx R_{\text{int}}/1000)$ revolved around the $Z-$axis.
This approach  leads to faster convergence and more accurate results than prescribing a membrane geometry. \\
We write the relaxed strain energy density as
\begin{subequations}
\begin{equation}
\begin{split}
&W^{\pazocal{R}}\left(\lambda_{R}, \lambda_{\Phi}, \lambda_{Z}, E, \nu\right) = \Big(\textsc{LogOpt}_{1}\left(\lambda_{R}, \lambda_{\Phi}, \nu\right)\Big) \left(\left(\frac{\Lambda + 2\mu}{2}\right)(I_{1})^2 - 2\mu I_{2}\right) \\
&+ \Big(\textsc{LogOpt}_{2}\left(\lambda_{R}, \lambda_{\Phi}, \nu\right) \Big)\left(\frac{E}{8}\left(\lambda_{R}^2-1\right)^2\right) + \Big(\textsc{LogOpt}_{3}\left(\lambda_{R}, \lambda_{\Phi}, \nu\right)\Big) \left(\frac{E}{8}\left(\lambda_{\Phi}^2-1\right)^2\right).
\end{split}
\label{eq:13a}
\end{equation}
where
\begin{equation}
\begin{split}
&\textsc{LogOpt}_{1}\left(\lambda_{R}, \lambda_{\Phi}, \nu\right) = \Big(\textsc{LogOpt}\big(\lambda_{\Phi} \geq \lambda_{\Phi}^{*}\left(\lambda_{R}, \nu\right) \big)\Big)\Big(\textsc{LogOpt}\big(\lambda_{R} \geq \lambda_{R}^{*}\left(\lambda_{\Phi}, \nu\right) \big)\Big) \\
& \textsc{LogOpt}_{2}\left(\lambda_{R}, \lambda_{\Phi}, \nu\right) =\Big(\textsc{LogOpt}\big(\lambda_{\Phi} \leq \lambda_{\Phi}^{*}\left(\lambda_{R}, \nu\right) \big)\Big)\Big(\textsc{LogOpt}\big(\lambda_{R} \geq 1 \big)\Big), \\
& 
\textsc{LogOpt}_{3}\left(\lambda_{R}, \lambda_{\Phi}, \nu\right)= \Big(\textsc{LogOpt}\big(\lambda_{R} \leq \lambda_{R}^{*}\left(\lambda_{\Phi}, \nu\right) \big)\Big)\Big(\textsc{LogOpt}\big(\lambda_{\Phi} \geq 1 \big)\Big), \\
&I_{1}=\trace{\mathbf{E}_{\pazocal{G}}}, \qquad I_{2}=\nicefrac{1}{2}\Big(\left(\trace{\mathbf{E}_{\pazocal{G}}}\right)^2 - \trace{(\mathbf{E}_{\pazocal{G}}^2)}\Big).
\end{split}
\label{eq:13b}
\end{equation}
\end{subequations}
Here, $I_{1}, I_{2}$ are two principal invariants of the Green-Lagrange strain tensor $\mathbf{E}_{\pazocal{G}} = (\mathbf F^T\mathbf{F}-\mathbf{I})/2$ and $\mathbf{I}$ is the Identity tensor. The first and second \text{Lamé}'s parameters are given by $\Lambda$ and $\mu$ respectively. The logical operator is given by $\textsc{LogOpt}(\cdot)$ and is similar to the Heaviside function in \cref{eq:11b} and results in a faster convergence in \texttt{COMSOL}. 
Note that $\textsc{LogOpt}\big(\text{True}\big) =1$ and $\textsc{LogOpt}\big(\text{False}\big) =0$. 

%%%%%%%%%%%%%%%%%%%%%%%%%

\section{Wrinkling regions: quantitative results for Case 1, 2, 3}
%%%%%%%%%%%%%%%%%%%%%%%%%
% The quantitative results for regions of wrinkling for different cases are tabulated below
\begin{table}[ht]
\centering
\caption{\label{tab:curwrinklvalueset1} Case 1: Wrinkling regions for different traction loads in the current configuration.}
\begin{tabular*}{\linewidth}{l @{\extracolsep{\fill}}ccccc}
\hline\hline
\textbf{Quadrant}                                                    & \textbf{\begin{tabular}[c]{@{}c@{}}Applied \\ Traction Load \\ (in MPa)\end{tabular}} & \textbf{\begin{tabular}[c]{@{}c@{}}Referential\\  Radius\\  (in cm)\end{tabular}} & \textbf{\begin{tabular}[c]{@{}c@{}}Current \\ Radius \\  (in cm)\end{tabular}} & \textbf{\begin{tabular}[c]{@{}c@{}}Wrinkling\\  Region \\ (in cm)\end{tabular}}        \\ \hline\hline
\textbf{\begin{tabular}[c]{@{}c@{}}Left \\ Top\end{tabular}}     & 0.15                                                                                  & R = {[}0.5, 4.0{]}                                                                & r = {[}0.5, 4.5259{]}                                                          & r = {[}0.5, 0.7214{]}                                                                  \\ \hline
\textbf{\begin{tabular}[c]{@{}c@{}}Right\\  Top\end{tabular}}    & 0.50                                                                                  & R = {[}0.5, 4.0{]}                                                                & r = {[}0.5, 5.3329{]}                                                          & \begin{tabular}[c]{@{}c@{}}r = {[}0.5, 0.8877{]}\\ $\cup$ {[}1.6706, 2.0294{]}\end{tabular} \\ \hline
\textbf{\begin{tabular}[c]{@{}c@{}}Right \\ Bottom\end{tabular}} & 0.75                                                                                  & R = {[}0.5, 4.0{]}                                                                & r = {[}0.5, 5.7573{]}                                                          & \begin{tabular}[c]{@{}c@{}}r = {[}0.5, 1.0049{]}\\ $\cup$ {[}1.5764, 2.3246{]}\end{tabular} \\ \hline
\textbf{\begin{tabular}[c]{@{}c@{}}Left \\ Bottom\end{tabular}}  & 1.50                                                                                  & R= {[}0.5, 4.0{]}                                                                 & r = {[}0.5, 6.7085{]}                                                          & r = {[}0.5, 2.8549{]}                                                                  \\ \hline
\end{tabular*}
\end{table}

%%%%%%%%%%%%%%%%%%%%%%%%%%
\begin{table}[ht]
\centering
\caption{\label{tab:curwrinklvalueset3} Case 2: Wrinkling regions for different traction loads in the current configuration.}
\begin{tabular*}{\linewidth}{l @{\extracolsep{\fill}}ccccc}
\hline\hline
\textbf{Quadrant}                                                    & \textbf{\begin{tabular}[c]{@{}c@{}}Applied \\ Traction Load \\ (in MPa)\end{tabular}} & \textbf{\begin{tabular}[c]{@{}c@{}}Referential\\  Radius\\  (in cm)\end{tabular}} & \textbf{\begin{tabular}[c]{@{}c@{}}Current \\ Radius \\  (in cm)\end{tabular}} & \textbf{\begin{tabular}[c]{@{}c@{}}Wrinkling\\  Region \\ (in cm)\end{tabular}}        \\ \hline\hline
\textbf{\begin{tabular}[c]{@{}c@{}}Left\end{tabular}}     & 0.25                                                                                  & R = {[}0.5, 4.0{]}                                                                & r = {[}0.5, 6.0563{]}                                                          & \begin{tabular}[c]{@{}c@{}}r = {[}0.5, 0.8205{]}\\ $\cup$ {[}2.2160, 2.6328{]}\\ $\cup$ {[}4.4591, 6.0563{]}\end{tabular}                                                                \\ \hline
\textbf{\begin{tabular}[c]{@{}c@{}}Right\end{tabular}}    & 2.50                                                                                  & R = {[}0.5, 4.0{]}                                                                & r = {[}0.5, 10.9584{]}                                                          & \begin{tabular}[c]{@{}c@{}}r = {[}0.5, 1.3481{]}\\ $\cup$ {[}3.5796, 4.2844{]}\\ $\cup$ {[}7.7125, 10.7820{]}\end{tabular} \\ \hline
\end{tabular*}
\end{table}

%%%%%%%%%%%%%%%%%%%%%%%%%

\begin{table}[ht]
\centering
\caption{\label{tab:curwrinklvaluempset2} Case 3: Wrinkling regions for different traction loads in the current configuration.}
\begin{tabular*}{\linewidth}{l @{\extracolsep{\fill}}ccccc}
\hline\hline
\textbf{Quadrant}                                                    & \textbf{\begin{tabular}[c]{@{}c@{}}Applied \\ Traction Load \\ (in MPa)\end{tabular}} & \textbf{\begin{tabular}[c]{@{}c@{}}Referential\\  Radius\\  (in cm)\end{tabular}} & \textbf{\begin{tabular}[c]{@{}c@{}}Current \\ Radius \\  (in cm)\end{tabular}} & \textbf{\begin{tabular}[c]{@{}c@{}}Wrinkling\\  Region \\ (in cm)\end{tabular}}        \\ \hline\hline
\textbf{\begin{tabular}[c]{@{}c@{}}Left\end{tabular}}     & 0.05                                                                                  & R = {[}0.5, 4.0{]}                                                                & r = {[}0.5, 4.6326{]}                                                          & \begin{tabular}[c]{@{}c@{}}r = {[}0.5, 0.9731{]}\\ $\cup$ {[}2.7947, 3.5341{]}\end{tabular}                                                                \\ \hline
\textbf{\begin{tabular}[c]{@{}c@{}}Right\end{tabular}}    & 0.20                                                                                  & R = {[}0.5, 4.0{]}                                                                & r = {[}0.5, 10.9584{]}                                                          & \begin{tabular}[c]{@{}c@{}}r = {[}0.5, 1.1554{]}\\ $\cup$ {[}2.7960, 3.5945{]}\end{tabular} \\ \hline
\end{tabular*}
\end{table}

%%%%%%%%%%%%%%%%%%


\end{document}